# Technology of the SoLid detector and construction of the first submodule

**Celine Moortgat**[1,2,3]

*Ghent University*
*Proeftuinstraat 86, 9000 Ghent, Belgium*
*E-mail:* `celine.moortgat@ugent.be`

As submitted:

The SoLid experiment aims to resolve the reactor anti neutrino anomaly by searching for short baseline neutrino oscillations. The experiment makes use of a novel detector technology based on the combination of 5cm×5cm×5cm PVT cubes and $^6$LiF:ZnS screens. This technology provides an improvement for the background rejection capabilities, the neutron identification and the localization of the inverse beta decay compared to the standard liquid scintillators + Gd detectors. These proceedings discuss the construction and commissioning of the first module installed at the BR2 research reactor in SCK-CEN, Mol, Belgium. Radioactive sources are used to study the light attenuation and to calibrate the energy response of the detector.

---

[1] Speaker
[2] Aspirant of the FWO-SCK.CEN
[3] For the SoLid Collaboration

# 1. Introduction

Neutrinos are the least understood particles in the present Standard Model of particle physics. They interact with matter exclusively through the weak interaction, which makes their detection very difficult. Within the Standard Model there are three flavours of neutrinos: the electron-, muon- and tau-neutrino. The Standard Model implies no interaction between the neutrinos and the recently confirmed scalar field, thus yielding massless neutrinos. Since the 1960s increasing experimental evidence showed the known neutrinos to be massive, which was unambiguously established by the observation of neutrino oscillations by the Super-Kamiokande [1] and SNO [2] experiments at the turn of the 21$^{st}$ century. This key observation opens a portal for the discovery of new physics phenomena that go beyond the Standard Model. In most elementary extensions of the Standard Model, the neutrino masses imply the existence of minimally one, but possible more non-interacting (sterile) right-handed neutrinos.

Most of the current world data is in agreement with the simple picture of the mixing of three neutrino flavours, except for a few striking anomalies. The LSND experiment at Los Alamos [3] measured an excess of electron anti-neutrinos and attributed it to the oscillation of low energy muon anti-neutrinos originating from muon decays at rest. Their data implied a large mass difference between two neutrino species which is not confirmed by more recent neutrino experiments. Performing a similar search, the MiniBooNe experiment at Fermilab [4] observed a similar excess of electron anti-neutrinos originating from a muon anti-neutrino beam, but got no clear confirmation of this result when looking for electron neutrinos using a muon neutrino beam. The current interpretation of the LSND and the MiniBooNE results requires the introduction of at least two more light sterile neutrinos with masses in the 1 eV range. Gallium detectors used for solar neutrino experiments, GALLEX [5] and SAGE [6], have observed 14 ± 5% less neutrinos than expected from radioactive sources placed inside the detector for calibration purposes. An interpretation in terms of an additional sterile neutrino is given in [7], where a remarkable compatibility between the MiniBooNE and Gallium data is demonstrated.

Recently, existing short baseline reactor electron anti-neutrino oscillation data have been re-analyzed [8] by using the results of a new and very detailed calculation of the reactor neutrino flux [9]. The improved flux calculations, which are on average 3% larger than former computations, are based on accurate predictions from ab-initio calculations of all beta emitting isotopes associated with the fission of $^{238}$U in combination with reference beta decay spectra measurements and nuclear database input from $^{235}$U, $^{239}$Pu and $^{241}$Pu isotope decays. The experimental result is expressed in the ratio between the observed and calculated electron-anti neutrino flux under the hypothesis of three active oscillating neutrinos which amounts to 0.943±0.023, leading to a deviation from unity at 98.6% C.L [8]. This significant deviation is now known as the reactor neutrino anomaly. Assuming the correctness of the new flux factors, the reactor anomaly is unlikely to be explained by a common bias in a wide variety of reactor experiments, based on different detection technologies and fuel mixtures. The other possible explanation of the anomaly is based on a real physical effect involving the existence of a fourth nonstandard neutrino, corresponding in the flavour basis to a sterile neutrino.

By performing experiments with efficient electron anti-neutrino detectors very close to the core of a nuclear fission reactor one can attempt to resolve the reactor anomaly. The international SoLid Collaboration will construct a competitive short baseline neutrino detector at the BR2 research reactor of the SCK in Mol. The detector can be operated at distances between 5.5 and 10 meters from the highly enriched uranium core. The reactor specifications are nearly ideal due to the small core size, stability and high power of operation. However, working in the environment close to a nuclear reactor still raises a number of experimental challenges. There are high rates of background events due to the low overburden and close proximity to the reactor.

To overcome these challenges, a novel detector technology was developed which is explained in section 2. The use of this new approach required the production and deployment of prototype detectors to demonstrate and understand the new technology. A 288 kg full scale detector module was deployed at the BR2 reactor in November 2014, which took reactor-on, reactor-off and calibration source data between February and September 2015. The 288 kg module and its deployment are detailed in section 3. The analysis of the data provided by this module is ongoing. Starting in 2016 the SoLid collaboration will perform a phased experiment that will provide world-leading sensitivity to oscillations within 10 m of a nuclear reactor.

## 2. SoLid anti neutrino detection technology

The detection of electron anti neutrinos is based on the inverse beta-decay (IBD) process where an electron antineutrino interacts with a proton from the detector to produce a positron and a neutron (figure 1). The SoLid detector is made of 5 cm Polyvinyl toluene (PVT) cubes.

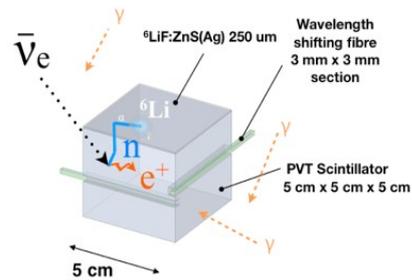

Figure 1: IBD process in a SoLid detector PVT cube.

The positron is detected by the scintillation light it generates in the interaction within PVT cube. Afterwards it annihilates with an electron in the detector, emitting a pair of 511 keV γ's. It is the energy of the positron that will be used to reconstruct the neutrino energy. The energy deposited in the interaction cube can be used as a good estimate for the positron energy, if one takes into account the small fraction of annihilation gamma energy also deposited in the same cube. The neutron generated in the IBD event is, after thermalisation in the cube, captured by the layer of $^6$Li:ZnS(Ag) and produces a tritium nucleus and an alpha particle which excite the surrounding grains of ZnS(Ag) scintillator. Due to the high deposited energy, the scintillator is put in an excited state that decays slowly, up to microseconds after the capture (figure 2). By the use of pulse shape discrimination this signal can be well distinguished from the positron signal.

The scintillation light generated in the cubes and the ZnS(Ag) screens is captured by a wavelength shifting (WLS) fibre and transported to the MPPC (multi-pixel photon counter) present on one end of the fibre. Because there is a fibre in the x- and the y-direction the original cube of the scintillation light can be determined. In this way, positron candidates and neutron captures can be located to within 5 cm. The advantage of this novel detection technique over the standard Gd doped liquid scintillator approach is that due to such a high segmentation the events can be imaged spatially. This is a very powerful technique in the discrimination of IBD events versus background. For example, for an IBD event the positron and neutron signal are expected to be close to each other, so that time correlated signals far from each other can be discarded as not being IBD events.

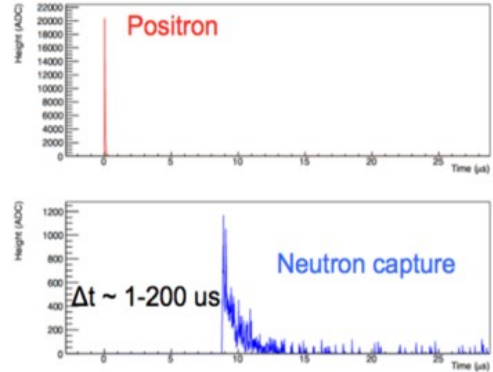

Figure 2: Comparison of scintillation signals coming from the positron ionization and neutron capture.

## 3. Construction of the first submodule

In the summer of 2014 a 288 kg module (figure 3) was constructed. The module consists out of 2304 5cm PVT cubes, each equipped with a $^6$LiF:ZnS(Ag) screen and wrapped in Tyvek. All individual parts were precisely measured to get an accurate determination of the total number of protons in the detector. This allowed for a precision on the proton number of better than 1%. The assembled cubes were arranged in 9 planes of 16x16 cubes.

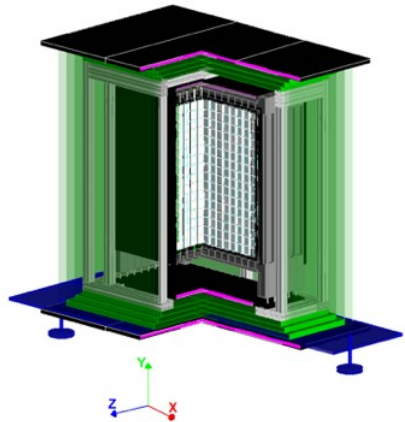

Figure 3: Diagram of the first detector sub module.

Each plane is individually supported by an aluminium frame and has a wavelength shifting fibre grid. The grid consists of 16 horizontal and 16 vertical fibres, read out at one end by a 3 mmx3 mm 50 micron pitch MPPC and equipped with a reflecting mirror on the other side. The coupling to the MPPC is performed with a custom 3D printed socket, shown in figure 4. The MPPCs are connected via a small PCB to a micocoax cable, which carries the bias voltage for and the signal from the MPPC to the detector electronics.

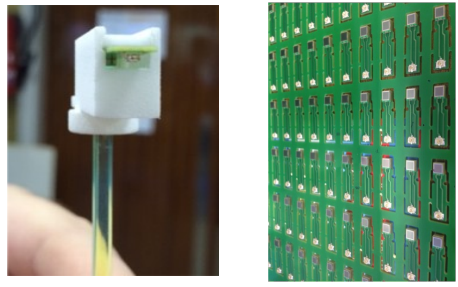

Figure 4: Left, fibre inserted into 3D printed socket. Right, sensor PCB production.

### 3.1 Electronics & trigger of SM1

The electronics for the first submodule consists of 5 modules, handling the 9 planes and the external muon vetos. Each module has a gigabit link interface board (GLIB) [10], a slow control board and 2 analog/digitizer boards. The latter provide an individually programmable bias voltage for each of the 32 connected MPPCs using an HV chip and trimmed using a multi-channel digital to analog converter, controlled by the slow control board. The signals are amplified and digitised with a 14 bit resolution at a sampling rate of 65 MHz using custom electronics. The signals coming from the 2 analog/digitizer boards are received by a single GLIB board which performs a per channel threshold based trigger. It requires at least one vertical fibre's channel and at least one horizontal fibre's channel to go above threshold within 3 time samples in order for a trigger to occur. When a trigger occurs, waveforms of 256 samples are read out from the channels above threshold in the plane where the trigger occurred. In addition, all channels are synchronously triggered by a Poisson distributed random trigger pulse. The data is read from the GLIBs using the IPbus v2 protocol [11] over a local gigabit Ethernet network. The slow control board measures the temperature given by sensors located on the analog/digitizer board and on the detector frames and accordingly sets the trim voltages for the MPPC bias voltages.

### 3.2 Quality assurance tests

Some quality assurance tests were performed to check the full readout chain of the individual channels. For this purpose each plane was placed horizontally under a dark cloth. The optics were tested by placing a $^{60}$Co source on top of some cubes. For every WLS fibre the source was placed on the cube that is situated at the opposite end from the MPPC. In this way, the amplitude spectrum was measured for each fibre using random triggers with a trigger threshold of 100ADC counts above pedestal. During this measurement it became clear that all channels were correctly connected and that all fibres, MPPCs and cables were intact. For the edge channels of each plane, measurements with the $^{60}$Co source were performed at 9 different cubes along the fibre so to have data from different distances from the MPPC. With these data a study of the light attenuation throughout the fibre could be performed.

In figure 5 the mean value of the maximum amplitudes measured at a certain distance from an MPPC for all edge channels are shown vs the distance along the fibre. In this way it was determined that the average amplitude reduction along a row of cubes is of the order of 15%.

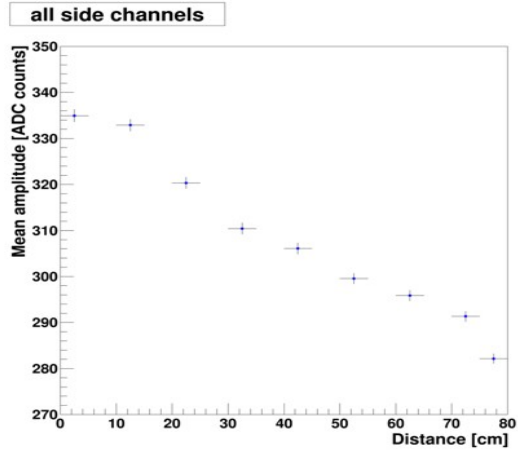

Figure 5: Mean amplitude of the scintillation signal vs. distance from the MPPC.

For some planes cosmic muon data was taken overnight with random triggers using a trigger threshold of 350 ADC counts above pedestal. As expected, the hits are uniformly distributed over each plane although the amount of hits is too limited to make a conclusive statement [12].

### 3.3 Commissioning and data taking

After the manufacturing of the individual parts, the 9 planes were assembled at Ghent University and put together to create the detector. On top of and below the detector muon vetos were installed. These PVT slabs are fitted with a WLS fibre coupled to MPPCs on both ends. Four muon veto panels are located on top and four are located below the detector. The detector as a whole is surrounded by a HDPE shield of 9 cm thickness to shield the detector from thermal neutrons and to reflect the neutrons coming from IBD events close to the detector's surface back into the detector. The detector was moved and installed at the BR2 reactor at the end of November 2014. The commissioning took place between December and mid-February. During this period the 302 MPPC channels were equalised to the same gain. After this, the threshold trigger scheme was commissioned to minimise the threshold whilst minimising the dead time of the DAQ system due to high instantaneous trigger rates.

The detector entered a stable physics run in mid-February 2015 when the reactor was running at approximately 70 MW. After this reactor-on period the reactor shut down for a long maintenance period. During this time the detector kept running in the same conditions for 2 more months to get a good estimate of the background present on site. Afterwards testing with radioactive sources was performed. A $^{60}$Co and an AmBe source were placed at different positions around the detector. Another test was performed using a $^{252}$Cf source embedded in a PVT cube in the centre of the detector. The radioactive source campaign was completed by the end of September 2015 and will be used in the determination of the neutron detection efficiency and the detector's energy scale and resolution. The analysis of the full data sets is ongoing, but the first preliminary results are becoming available [13], [14], [15].

## 4. Conclusion

The SoLid Collaboration intends to resolve the reactor neutrino anomaly by exploring a novel composite scintillator and highly segmented detector technique. A 288 kg module has already been constructed and commissioned at the BR2 reactor in SCK-CEN, Belgium. This sub module took data during a reactor-on and reactor-off periods. In addition a number of calibration source tests were performed. This data is currently being analysed. By the second half of 2016 the next phase of the experiment starts and the first modules of the 3 tonne detector will be deployed at the reactor site.